\documentstyle[draft,11pt,psfig]{nature}

\title{Magnetar-like X-ray Bursts from an Anomalous X-ray Pulsar}
\mainauthor{}

\author{F. P. Gavriil\affiliation[1]{Physics Department, McGill University,
Montreal, QC H3W 2C4, Canada},
V. M. Kaspi$^{\ast}$\affiliation[2]{MIT Department of Physics and 
Center for Space Research, Cambridge, MA 02139, USA},
P. M. Woods\affiliation[3]{Space Science Research Center, National Space Science
and Technology Center, Huntsville, AL 35805, USA; Universities Space Research Association}
}

\begin{document}
\summary{
Anomalous X-ray Pulsars (AXPs) are a class of rare X-ray pulsars whose
energy source has been perplexing for some 20
years\cite{fg81,vtv95,ms95}.  Unlike other, better understood X-ray
pulsars, AXPs cannot be powered by rotation or by
accretion from a binary companion, hence the designation ``anomalous.''
AXP rotational and radiative properties are strikingly similar to those
of another class of exotic objects, the Soft Gamma Repeaters (SGRs).
However, the defining property of SGRs, namely their low-energy
gamma-ray and X-ray bursts, have heretofore not been seen in AXPs.
SGRs are thought to be ``magnetars,'' young neutron stars powered by
the decay of an ultra-high magnetic field\cite{td95,kds+98}.  The
suggestion that AXPs are magnetars has been
controversial\cite{chn00}.  Here we report the discovery, from the
direction of AXP 1E~1048$-$5937, of two X-ray bursts that have many
properties similar to those of SGR bursts.  These events imply
a close relationship between AXPs and SGRs, with
both being magnetars.  }

\maketitle

SGRs are believed to be magnetars because the high magnetic field
provides the torque for their rapid spin-down, as well as 
the energy to power their bursts and quiescent X-ray
emission\cite{td95}.  AXPs have been suggested to be magnetars,
albeit less active, because of their similar spin periods, rates of
spin down, location in the Galactic plane, and similar though somewhat
softer X-ray spectra to those of SGRs in quiesence\cite{td96a}.  The
physical difference between the two classes is unknown, but, in the
magnetar model, is likely related to the magnitude or distribution of
the stellar magnetic field.  However, the apparent absence of any
bursting behavior in AXPs has led to suggestions that they could be
powered, not by magnetism, but by accretion from a disk of material
remaining after the birth supernova event\cite{chn00}.  If so, the
observational similarities between AXPs and SGRs must be purely
coincidental.

A program to regularly monitor the AXPs using the Proportional Counter Array
(PCA)\cite{jsg+96} aboard NASA's {\it Rossi X-ray Timing Explorer
(RXTE)} was begun in 1996 in order to determine their long-term timing,
pulsed flux, and pulse profile stabilities\cite{kcs99,klc00,kgc+01,gk02}.  As
part of this program, motivated by the existence of SGR bursts, we also
searched the AXP data for bursts (see Fig. 1 caption for details).

We discovered two highly significant bursts from the direction of
AXP 1E~1048$-$5937 in this way.  The first (hereafter Burst 1) occurred
during a 3-ks PCA observation obtained on 2001 October 29 with chance
probability $P \simeq 6\times10^{-18}$ after accounting for the number
of trials. A second burst (hereafter Burst 2) was found in a 3-ks
observation obtained on 2001 November 14, with analogous 
probability $P \simeq 2\times10^{-9}$.  No other
significant bursts were found toward 1E~1048$-$5937.  The total PCA
time searched for bursts toward this source was 380~ks in observations
obtained from 1996-2002.  

\begin{center}
\begin{figure}
\centerline{\psfig{file=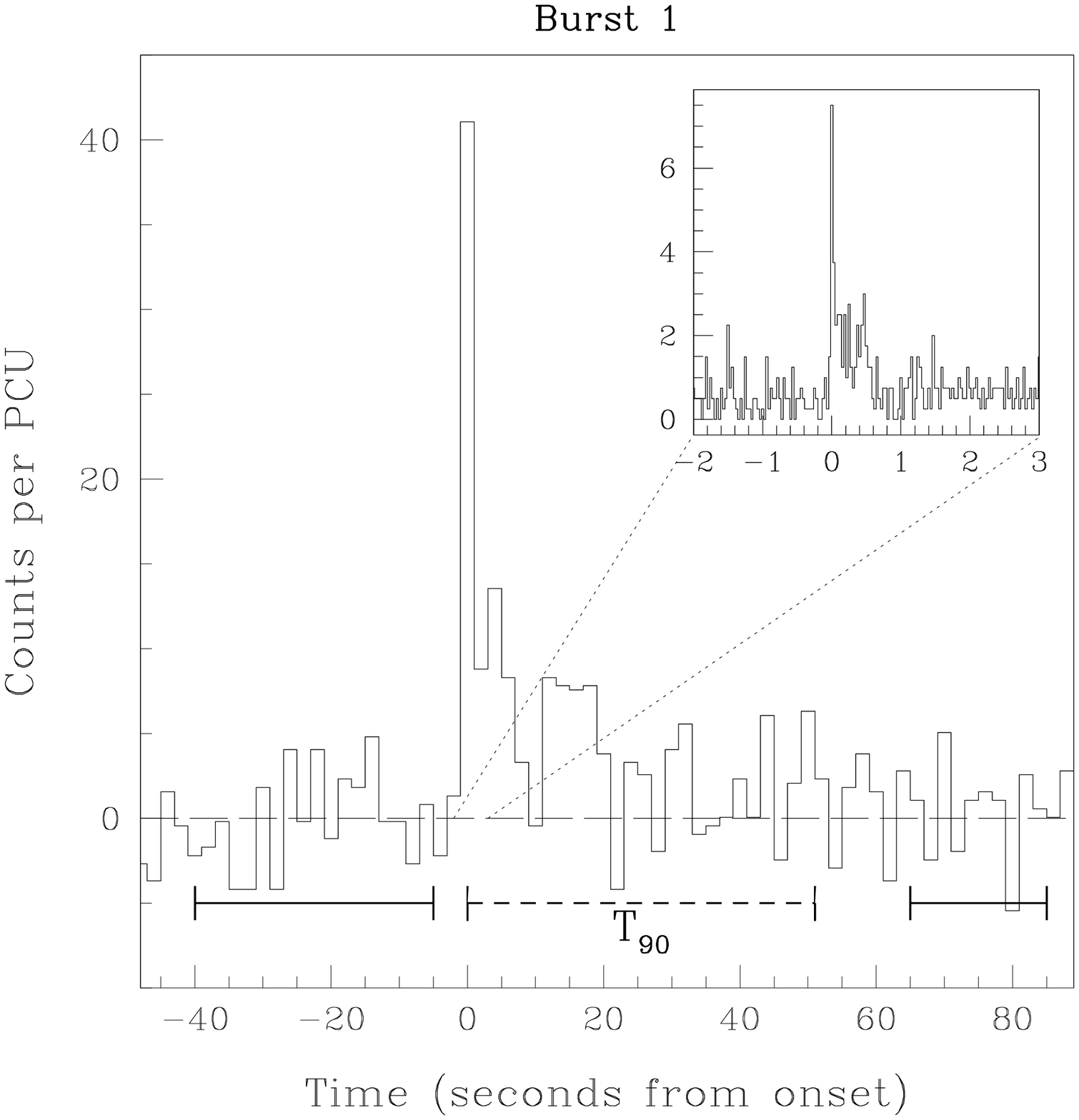,height=9cm}\psfig{file=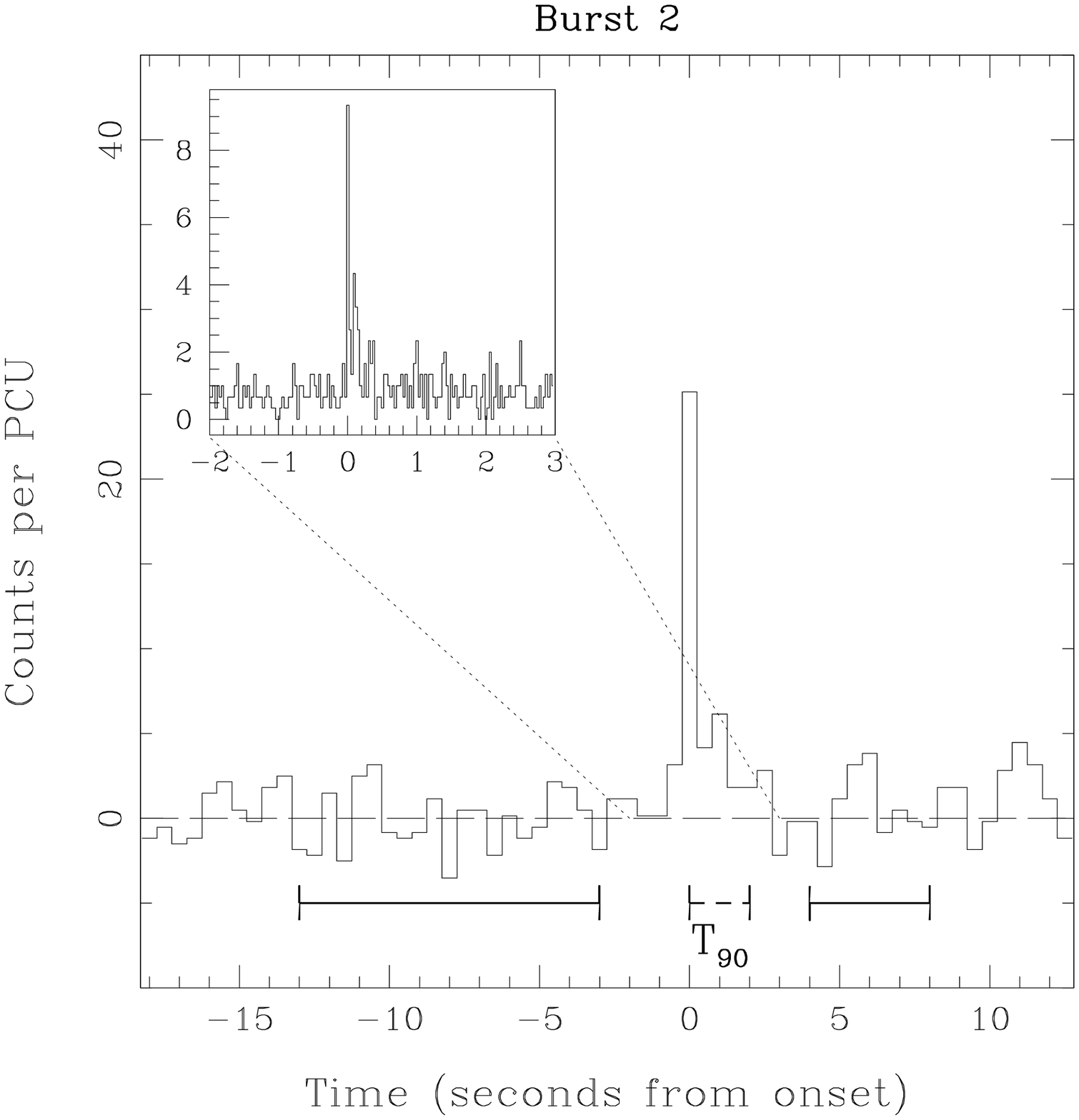,height=9cm}}
\caption{}
\label{fig:profiles}
\end{figure}
\end{center}

The burst profiles are shown in Figure~\ref{fig:profiles}.  Both are
characterized by fast rises and slow decays (see Table~1).  Burst 1
appears to have a long, low-level tail that is just above the PCA
background as determined by intervals selected before and after the
bursts (see Fig.~\ref{fig:profiles}), while Burst 2 is much shorter.
Both bursts arrived at the peak of the AXP pulse within uncertainties
in burst arrival time and definition of pulse peak.  The probability of
this occurring by random chance is $\sim$1\%.  We note a marginal
($\sim 3\sigma$) increase in the pulsed flux from 1E~1048.1$-$5937 that
commenced with the observation in which Burst~1 was detected, and which
lasted $\sim$4~weeks.

To determine the bursts' spectral properties, 
we first established that neither burst exhibited significant spectral
evolution with time by computing hardness ratios (the ratio of
10--60~keV counts to 2--10~keV counts) for the first 0.5-s and
subsequent 1.5-s burst intervals.  No significant change in hardness
was detected, though marginal spectral softening with time was detected
after the first 2.5~s of Burst 1.  Hardness ratios for Burst 1 and 2
for the 1~s after burst onset were $2.8 \pm 0.8$ and $1.0 \pm 0.3$,
respectively.  

\begin{center}
\begin{figure}
\centerline{\psfig{file=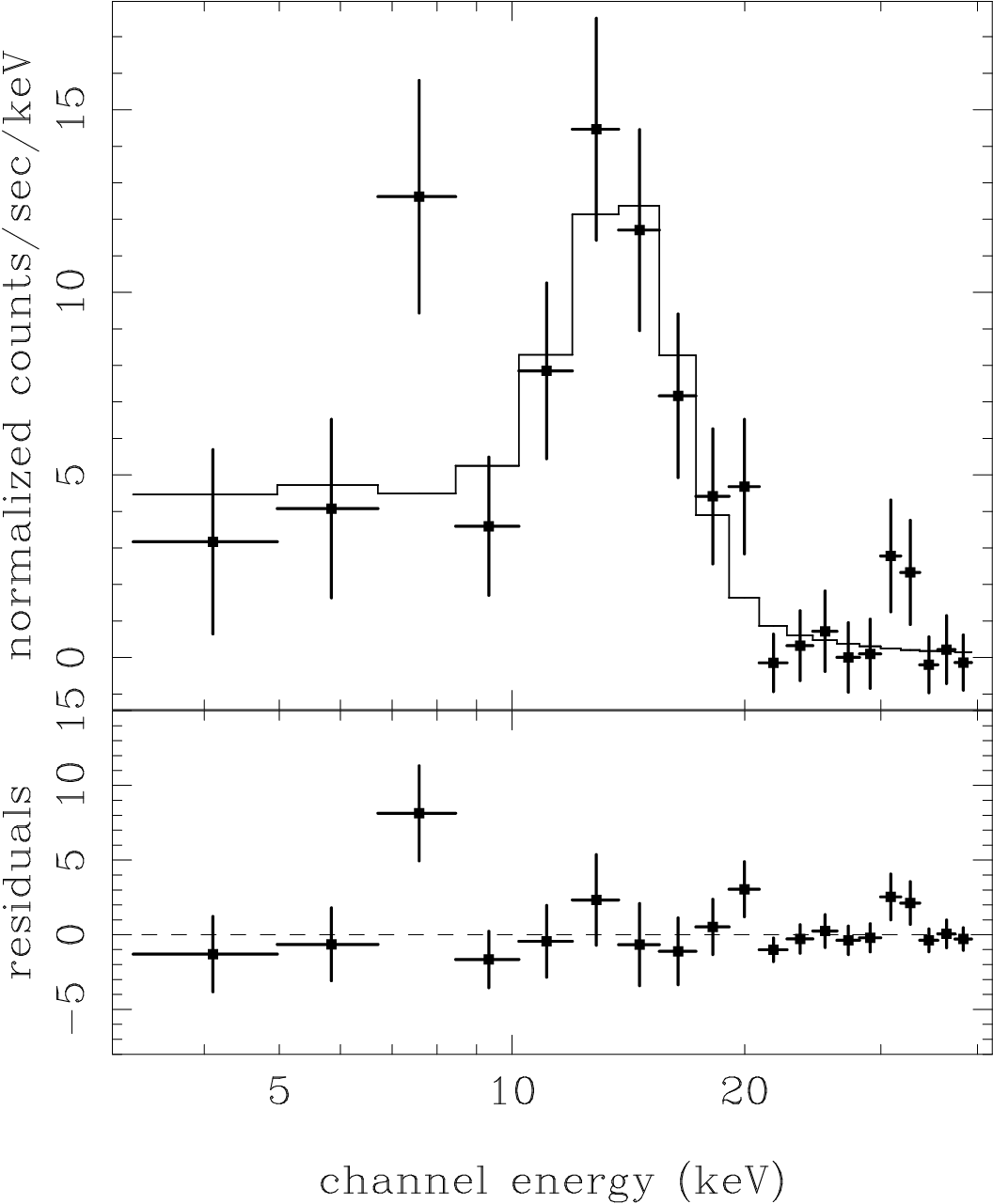,height=15cm}}
\caption{}
\label{fig:b1spectra}
\end{figure}
\end{center}

We then fit the spectra from the first 1~s of each burst to two
one-component models, a power law and a black body (see Table 1).
Continuum models provided an adequate characterization of the Burst~2
spectrum but not of the Burst~1 spectrum.  As seen in
Figure~\ref{fig:b1spectra}, the spectrum for the 1~s after the Burst 1
onset exhibits a feature near 14~keV.  This feature is clear
in all binning schemes and is prominent throughout the first
$\sim$1~s of the burst.  No known PCA instrumental effect produces a
feature at this energy (K. Jahoda, personal communication).

Due to the wide ($\sim$1$^{\circ}$) field-of-view (FOV) and lack of
imaging capabilities of the PCA, we cannot verify that the bursts
originated from the location of the AXP.  The low peak X-ray
fluxes of the events (see Table 1) preclude determining the source's
location using data from other, better imaging instruments that were
contemporaneously observing the X-ray sky, such as the {\it RXTE} All
Sky Monitor, or the Wide Field Camera aboard {\it BeppoSAX}.
We must therefore consider other possible origins from
the bursts before concluding they were from the AXP.

The bursts' short rise times (Table~1) require emission regions of less
than a few thousand km, implying a compact object origin.  So-called
Type I X-ray bursts are a well-studied phenomenon that result from
unstable helium burning just below the surface of a weakly magnetized
neutron star that is accreting material in a low-mass X-ray binary
(LMXB)\cite{lpt95}.  However, Type I bursts from an LMXB in the same
FOV as 1E~1048.1$-$5937 are unlikely to explain our observed
bursts because (i) the burst rise times are much shorter than those of
Type I bursts; (ii) the burst spectra are much harder than those of
Type I bursts; (iii) Burst 2 shows no evidence for spectral softening
with time and no Type I burst has ever exhibited a spectral feature
like the one detected in Burst 1; (iv) the bursts are extremely faint,
implying a source location well outside the Milky Way for Type
I burst luminosities (v) there are no known LMXBs in the
FOV\cite{lph01}.  Type II X-ray bursts\cite{lpt95} are a much rarer and
less well understood phenomenon observed thus far in only two sources,
both accreting binaries.  The bursts we have observed are unlikely to
be Type II bursts from an unknown X-ray binary in the PCA FOV because (i)
of the rarity of such events; (ii) Type II bursts have longer rise
times than do our bursts; (iii) no Type II burst has exhibited a
spectral feature like that seen in Burst 1.

Classical gamma-ray bursts (GRBs) sometimes exhibit prompt X-ray
emission that can have temporal and spectral signatures similar to
those we have observed\cite{hzkw01}.  However, the likelihood of two
GRBs occurring within $1^{\circ}$ of each other is small, and GRBs are
not known to repeat.  Conservatively assuming GRB spectral model
parameters that result in low gamma-ray fluxes and extrapolating the
GRB rate\cite{stk+01} as measured with the Burst and Transient
Source Experiment (BATSE)\cite{fmw+93} assuming homogeneity below the
BATSE threshold, we estimate a probability that these events are
unrelated GRBs that occured by chance in the same {\it RXTE} FOV during
our 1E~1048.1$-$5937 monitoring observations (conservatively neglecting
that they occured within two weeks of each other) of $\sim 9 \times
10^{-5}$.

The observed burst properties are in many ways similar to those seen
from SGRs\cite{gkw+01}.  The fast rise and slow decay profiles are
consistent with SGR time histories, as are the burst durations
(neglecting the long, low-level tail of Burst~1).  Both AXP and SGR
bursts are spectrally much harder than is their quiescent pulsed
emission.  The burst peak fluxes and fluences fall within the range
seen for SGRs, and the spectrum of Burst 2 is consistent with SGR burst
spectra of comparable fluence.  Burst 1 has characteristics unlike
nearly all SGR bursts, specifically its long tail and spectral
feature.  However, we note that a single event from SGR~1900+14 was
shown\cite{isw+01,si00} to possess each of these properties.  The
marginal increase in the pulsed fraction that we observed at the burst
epochs is consistent with SGR pulsed flux increases seen during bursting
episodes\cite{wkg+01}.  Finally, the fact that in spite of several years
of monitoring, the only two bursts detected occurred within two weeks of
each other suggests episodic bursting activity, the hallmark of SGRs.
Thus, the characteristics of these events match the burst properties of
SGRs far better than any other known burst phenomenon.

In the magnetar model for SGRs\cite{td95,td96a}, bursts are a result of
sudden crustal yields due to stress from the outward diffusion of the
huge internal magnetic field.  Such yields cause crust shears which
twist the external magnetic field, releasing energy.  Thompson \&
Duncan\cite{td96a} who, upon suggesting that AXPs are also magnetars,
predicted X-ray bursts should eventually be seen from them.  By
contrast, in no AXP accretion scenario, whether binary or isolated
fall-back disk, are SGR-like bursts expected.

The large 14-keV spectral line in Burst 1 is intriguing.  An electron
cyclotron feature at this energy $E$ implies a magnetic field of $B =
2\pi m c E / h e \simeq 1.2 \times 10^{12}$~G (where $m$ is the
electron mass, $c$ is the speed of light, $h$ is Planck's constant, and
$e$ is the electron charge), while a proton cyclotron feature implies
$B \simeq 2.4 \times 10^{15}$~G.  The former is significantly lower
than is implied from the source's spin-down and is typical of
conventional young neutron stars, rather than magnetars.  The latter is
higher than is implied by the spin-down yet reasonable for the magnetar
model as the spin-down torque is sensitive only to the dipolar
component of the magnetic field.

Why do the burst rates of AXPs and SGRs differ so markedly,
in spite of their common magnetar nature?  One possibility is
that AXP internal magnetic fields are much larger than those of SGRs;
if so, AXP crusts can undergo plastic deformation rather than brittle
fracturing\cite{td96a}.  However, this is opposite to what is inferred
from the two classes' spin-down rates, suggesting the latter is an
unreliable internal field indicator.  This could help reconcile the
contrasting radiative properties of AXPs and apparently high-magnetic
field radio pulsars\cite{pkc00}.  It also suggests that AXPs are SGR
progenitors, with bursting behavior commencing as the field decays.
This is consistent with the smaller AXP ages implied by their more
numerous associations with supernova remnants\cite{gvt99}, but does not
explain why AXPs and SGRs have similar spin period distributions, since
AXPs spin down as they age\cite{gsgv01}.  This aspect of magnetar
physics remains a puzzle.

%\bibliographystyle{nature}
%\bibliography{journals1,modrefs,psrrefs,crossrefs}

\newpage

\section*{Acknowledgements}
\noindent

We thank E. Kuulkers, K. Hurley, K. Jahoda, C.~Kouveliotou, W. Lewin,
M.~Lyutikov, M. Muno, D. Psaltis, S.~Ransom, M.~Roberts, D.~Smith, and
C.~Thompson for useful discussions.  This work was supported in part by
a NASA LTSA grant, an NSERC Research Grant, and an NSF Career Award.
VMK is a Canada Research Chair and an Alfred P. Sloan Fellow.  This
research has made use of data obtained through the High Energy
Astrophysics Science Archive Research Center Online Service, provided
by the NASA/Goddard Space Flight Center.

\bigskip
\bigskip
Correspondence and requests for materials should be addressed to
vkaspi@physics.mcgill.ca.

\newpage
\begin{center}
\begin{tabular}{lcc} \hline \hline
\multicolumn{3}{c}{Table 1\hspace{.5in} AXP Burst Timing and Spectral Properties} \\\hline
 &  Burst 1  & Burst 2 \\\hline
\multicolumn{3}{c}{\bf Temporal Properties} \\
Burst day, (MJD) & 52211 &  52227\\
Burst start time, (fraction of day, UT) & 0.2301949(24) & 0.836323379(68) \\
Burst rise time, $t_r$ (ms) & $21^{+9}_{-5}$ & $5.9^{+2.0}_{-1.2}$ \\
Burst duration, $T_{90}$ (s) &  $51^{+28}_{-19}$ & $2.0^{+4.9}_{-0.7}$\\
Burst phase & $-$0.018 $\pm$ 0.034  & 0.051 $\pm$ 0.032  \\
\multicolumn{3}{c}{\bf Fluxes and Fluences} \\
$T_{90}$ fluence (counts) & 485 $\pm$ 118 & 101 $\pm$ 15 \\
$T_{90}$ fluence ($\times 10^{-10}~\mathrm{erg~cm}^{-2}$) & 20.3 $\pm$ 4.8  & 5.3 $\pm$ 1.2 \\
1-s fluence  (counts)  & 117 $\pm$ 13  & 69 $\pm$ 10 \\
1-s fluence ($\times 10^{-10}~\mathrm{erg~cm}^{-2}$) & $5.9_{-1.9}^{+8.6}$  & $4.0_{-0.8}^{+3.5}$ \\
Peak flux for 64~ms ($\times 10^{-10}~\mathrm{erg~s^{-1}~cm}^{-2}$) & $31^{+45}_{-10}$ & $26^{+23}_{-5}$ \\
Peak flux for $t_r$~ms ($\times 10^{-10}~\mathrm{erg~s^{-1}~cm}^{-2}$)  & $54^{+79}_{-17}$ & $114^{+100}_{-23}$ \\
\multicolumn{3}{c}{\bf Spectral Properties} \\
{\bf Power law:} & & \\
power law index & $0.89^{+1.8}_{-0.71}$   & $1.38^{+0.75}_{-0.62}$ \\
power law flux ($\times 10^{-10}~\mathrm{erg~s^{-1}~cm^{-2}}$)  & $2.0_{-1.8}^{+8.4}$ & $4.0^{+3.5}_{-0.8}$ \\
line energy (keV) & $13.9 \pm 0.9$ & ... \\
line width, $\sigma$ (keV) & $2.2^{+1.3}_{-1.0}$ & ... \\
line flux ($\times 10^{-10}~\mathrm{erg~s^{-1}~cm^{-2}}$) & $3.9_{-1.6}^{+2.2}$ & ... \\
reduced $\chi^2$/degrees of freedom & 1.24/15 & 0.77/5 \\
{\bf Black body:} & & \\
$kT$ (keV) & $3.9^{+3.7}_{-2.7}$  & $3.6^{+2.2}_{-1.3}$ \\
black body flux ($\times 10^{-10}~\mathrm{erg~s^{-1}~cm^{-2}}$) & $2.4_{-2.1}^{+5.0}$   & $3.8_{-1.5}^{+3.3}$ \\
line energy (keV) & $14.2^{+1.1}_{-1.2}$ & ... \\
line width (keV) & $2.1^{+1.5}_{-1.3}$ & ... \\
line flux ($\times 10^{-10}~\mathrm{erg~s^{-1}~cm^{-2}}$) & $3.7_{-1.9}^{+2.2}$ & ... \\
reduced $\chi^2$/degrees of freedom & 1.23/15 & 1.66/5 \\
\hline
\end{tabular}
\end{center}

\newpage
\section*{Table Caption}
\noindent

{\bf Table 1.} 
The uncertainty on the burst start time is the burst rise time $t_r$
and is given in parenthesis as the uncertainty in the last digits
shown.  The burst rise times were determined by a maximum likelihood
fit to the unbinned data using a piecewise function having a linear
rise and exponential decay.  The burst duration, $T_{90}$, is the
interval between when 5\% and 95\% of the total 2--20~keV burst fluence
was received.  The background regions used for both the duration and
spectral analyses are shown in Figure~1.  Burst phase is defined such
that the peak of the periodic pulsation is at phase 0/1.  All fluences
and fluxes are in the 2--20~keV range.  $T_{90}$ fluences in cgs units
were calculated assuming a power-law spectral model and spectral grouping
that demanded a minimum of 20 counts per spectral bin.  The 1-s fluences
in cgs units correspond to the fluxes found in the spectral modeling.
The spectral rebinning method used in all spectral modeling for Burst 1
was to group the 256 PCA channels by a factor of 4, while  
for Burst 2, we demanded at least 20 counts per spectral bin.
Peak fluxes on the short time scales were determined by scaling
the 1-s fluxes by number of counts.
For all spectral fits, the equivalent neutral
hydrogen column density was held fixed at $1.2\times10^{22}$~cm$^{-2}$,
the value determined from recent XMM observations\protect\cite{tgsm02}.
Spectral fits were determined for the first 1~s after the burst start
times.  Spectral modeling was done using photons in the 2--40~keV
range.  
Response matrices were created using the {\tt FTOOL pcarsp}
\footnote{http://heasarc.gsfc.nasa.gov/docs/xte/recipes/pca\_response.html}.
Uncertainties in the Table are 68\% confidence intervals, except for
those reported for the cgs-unit fluences and fluxes, as well as the
spectral model parameters, for which we report 90\% confidence
intervals. 

\newpage

\newpage
\noindent
{\bf Figure 1 Caption}

\noindent
Lightcurves for the observed bursts.  The {\it RXTE} AXP data set consists
of short ($\sim$3~ks) snapshots, as well as longer archival
observations, all taken in the PCA {\tt GoodXenonwithPropane} mode,
which records photon arrival times with 1-$\mu$s resolution, and bins
photon energies into 256 channels.  Time series were initially created
with 31.25-ms resolution from photons having energies in the range
2--20~keV for each PCA Proportional Counter Unit (PCU) separately,
using all xenon layers.  Photon arrival times at each epoch were
adjusted to the solar system barycentre.  The resulting time series
were searched for significant excursions from the mean count rate by
comparing each time bin value with a windowed 7-s running mean.  Bursts
were identified assuming Poissonian statistics, and by combining
probabilities from the separate PCUs.  Left Panel:
Background subtracted 2-20~keV light curves for Burst 1, displayed with
2-s time resolution.  The solid horizontal lines before and after the
bursts are the boundaries of the pre- and post-background intervals
used for calculating $T_{90}$ and for spectral modeling.  The $T_{90}$
interval is shown as a horizontal dashed line.  Right Panel: Same but
for Burst 2, and with 0.5-s time resolution in the main panel.  The
insets show the peak of each burst with 31.25-ms time resolution.  We
verified that there was no significantly enhanced signal from PCA
events not flagged as ``good'' at the times of the bursts (such as
those that do not enter through the PCA aperture) in the {\it RXTE}
``Standard 1'' event files.  We also verified that both events were
clearly detected in all operational PCUs.  Hence the events are
unlikely to be instrumental in origin.

\newpage

\newpage
\noindent
{\bf Figure 2 Caption}

\noindent
X-ray spectrum in the 2--40~keV range for the 1~s after the onset of
Burst 1.  For all spectral analyses, we first created spectral files 
having 256 channels across
the full PCA energy range ($\sim$0.2--60~keV), although subsequent
factor of 4 rebinning was necessary because of the paucity of burst
counts.  The burst and background intervals were used as input to the
X-ray spectral fitting package {\tt XSPEC\cite{arn96}}
v11.1.0\footnote{http://xspec.gsfc.nasa.gov}.  The spectrum of the
first 1~s after Burst 1 onset is not well
characterized by any continuum model. The best fit power-law plus
Gaussian line model is shown as a solid line.  The F-test shows that
the addition of a line of arbitrary energy, width and normalization to
a simple power law model improves the fit significantly, with a chance
probability of this occuring of $0.0032$.  Monte Carlo simulations in
{\tt XSPEC} were done to verify this conclusion: 10,000 simulations of
similar data sets were produced assuming a simple power-law energy
distribution, then fit with a power law plus Gaussian line of arbitrary
energy, width and normalization.  This procedure is conservative, since
it ignores that the observed large line has flux comparable to the
measured continuum.  In 10,000 trials, we found 1 trial with the same
or smaller chance occurence probability as judged by the F-test,
indicating that the probability of the line we observed being due to
random chance is $<0.0001$.  We repeated this procedure for data having
a black-body spectrum, with similar results, namely the probability of
the line being due to chance is $<0.0008$.  The spectrum also shows
possible additional features at $\sim$7~keV and $\sim$30~keV
(suggestive of of lines at multiples of 1, 2 and 4 of $\sim$7~keV).
These additional features are not apparent in all binning schemes and
are not statistically significant.

\end{document}